# SLO-Scaler: Uncertainty-Aware SLO-Driven Autoscaling for Microservices

Shuo Wang
Carnegie Mellon University
Pittsburgh, USA

Xiaoxuan Sun
Independent Researcher
Mountain View, USA

Shao-yu Huang
Duke University
Durham, USA

Bencheng Su
University of Cincinnati
Cincinnati, USA

Shuo Xu
University of Southern California
Los Angeles, USA

Netra Awate*
Northeastern University
Boston, USA

***Abstract*—Autoscaling microservice-based applications to satisfy Service Level Objectives (SLOs) remains challenging due to bursty workloads, cascading latency across service dependencies, and cold-start overhead. Existing approaches such as the Kubernetes Horizontal Pod Autoscaler (HPA) rely on threshold-based CPU or memory metrics, which react too slowly to traffic spikes. Recent predictive methods improve responsiveness but generate point forecasts that ignore prediction uncertainty, leading to over-provisioning or oscillatory scaling. We propose SLO-Scaler, an uncertainty-aware autoscaling framework that predicts short-horizon request rates, tail latency, and SLO violation probability using a Bayesian LSTM model. SLO-Scaler integrates confidence-interval-based scaling decisions with a dependency graph analysis module that localizes bottleneck services, avoiding unnecessary whole-chain scaling. We evaluate SLO-Scaler on the DeathStarBench Social Network benchmark deployed on Kubernetes under periodic, bursty, and long-tail traffic patterns. Under bursty traffic, SLO-Scaler reduces the SLO violation rate by 29–56%, lowers the average replica count by 18–33%, and decreases scaling event frequency by 38–59% compared with the baselines, while achieving lower tail latency.**

## I. Introduction

Microservice architectures have become the dominant paradigm for building scalable cloud applications. Major cloud providers decompose monolithic applications into independently deployable services, each handling a specific business function [1]. While this architecture improves development agility and fault isolation, it introduces substantial complexity in resource management. Each microservice must be individually provisioned and scaled to meet end-to-end performance targets, typically expressed as SLOs on tail latency such as the 95th or 99th percentile response time.

Kubernetes has emerged as the de facto orchestration platform for containerized microservices. Its native HPA adjusts replica counts based on observed CPU or memory utilization against static thresholds. However, HPA exhibits several well-documented limitations [2]: threshold-based reactive scaling introduces a feedback delay of 30–90 seconds during which bursty traffic may cause SLO violations; CPU-centric metrics do not capture application-level indicators such as request latency; and HPA treats each service independently, ignoring inter-service dependencies that cause cascading latency amplification [3].

Recent research has explored predictive and reinforcement learning approaches. Predictive autoscalers use LSTM or GRU networks to forecast future workload, while RL methods learn scaling policies through environment interaction. However, these approaches share two shortcomings: most predictive models produce point estimates without quantifying uncertainty, risking over- or under-provisioning during workload regime changes; and existing methods typically scale all services uniformly without identifying which service is the actual bottleneck.

We propose **SLO-Scaler**, an uncertainty-aware autoscaling framework that addresses both limitations with three contributions:

1) A Bayesian LSTM prediction module that forecasts request rates, tail latency, and SLO violation probability with estimated confidence intervals.
2) An uncertainty-constrained scaling decision engine that uses predicted violation probability and confidence bounds to reduce false-positive scaling actions.
3) A bottleneck localization module based on dependency graph analysis that identifies performance-limiting services for targeted scaling.

We evaluate SLO-Scaler on the DeathStarBench Social Network benchmark deployed on Kubernetes, using the Alibaba 2022 cluster trace for workload generation.

## II. Related Work

Kubernetes HPA and KEDA (Kubernetes Event-Driven Autoscaler) are widely deployed but fundamentally reactive [4]. To overcome their latency, predictive methods use time-series forecasting: pHPA employs LSTM to predict request rates [5]; Graph-PHPA combines LSTM with GNNs for spatial dependencies [6]; Sinan predicts end-to-end latency with ML [7]; and MicroScaler uses online learning [8]. However, these methods produce point forecasts without uncertainty quantification.

RL-based autoscalers learn policies through trial-and-error [9], including meta-learning approaches for SLO-aware resource allocation [10] and holistic resource management [11], but require substantial training time and exhibit instability during exploration. DeepScaling considers service dependencies for holistic scaling decisions [12]. Rossi et al. investigated Bayesian deep learning for uncertainty-aware workload prediction [13]. Our work differs by jointly integrating uncertainty-aware SLO violation prediction with bottleneck-aware dependency graph analysis.

*Corresponding author.

Predictive autoscaling increasingly combines workload forecasting with system-specific constraints rather than treating resource demand as an isolated time series. SLO-aware graph forecasting directly links API and microservice dependency structure to scaling objectives [14], while attention-enhanced long-sequence modeling targets CPU-load dynamics over extended horizons [15]. Cold-start budgeting for serverless inference further couples forecasts with model-state tiering so that capacity decisions reflect startup cost as well as demand [16]. Together, these approaches strengthen anticipation and cost awareness, but they do not by themselves provide calibrated confidence bounds for deciding when a forecast is reliable enough to trigger scaling.

Uncertainty becomes most consequential under rare events, distribution shift, and shared infrastructure. Target-aware augmentation addresses rare-event prediction under tabular covariate shift [17], and calibrated risk scoring with structural regularization links representation quality to decision confidence [18]. Uncertainty modeling combined with external memory supports adaptation when learned representations encounter unfamiliar conditions [19], while risk-calibrated edge-cloud scheduling turns uncertainty estimates into reliability constraints for multi-tenant inference [20]. These lines of work motivate an autoscaling policy that treats predictive variance as a control signal rather than reporting it only as an auxiliary model statistic.

Resource coordination must also account for topology, heterogeneity, and avoidable communication. Graph-temporal representation learning captures evolving dependencies in dynamic risk identification [21]; drift-aware split federated learning adjusts cut layers and server updates to mitigate stragglers [22]; and learned client routing with aggregation selection optimizes cost across clouds [23]. At the cluster level, interference-aware coordination of training and inference uses learned models to characterize contention among heterogeneous GPUs [24]. Complementary efficiency mechanisms reduce the adaptation or collaboration footprint through subspace-deconvolution parameterization [25] and dynamic redundancy elimination in multi-agent systems [26]. These methods collectively point toward selective, dependency-aware actions in place of uniform scaling.

Reliable scaling depends on distinguishing a genuine bottleneck from transient symptoms. Retrieval-augmented fusion of logs, metrics, and traces builds a diagnostic twin for backend failures [27], while topology-aware localization under limited observability [28] and evidence-verified root-cause analysis [29] preserve causal structure when telemetry is incomplete. Evidence verification also constrains automated incident remediation [30] and tool execution [31], reducing the risk that a control action is based on unsupported diagnoses. This progression from multimodal telemetry to verified action complements SLO-Scaler's dependency-graph localizer by emphasizing traceable evidence before intervention.

Adaptation remains fragile when the data and knowledge supporting the controller drift over time. Lineage-aware semantic validation enables verified repair in data pipelines [32], and LLM-guided schema-drift repair operationalizes that principle for real-time ETL [33]. Experience-retrieval compression retains high-value adaptation evidence without unbounded memory growth [34], while process-aware multimodal analytics shows how temporally ordered evidence can be combined without discarding provenance [35]. These perspectives support maintaining auditable feature histories and compact experience summaries as the observability stream evolves, thereby separating durable changes from short-lived noise.

## III. System Design

### A. Overview

Fig. 1 illustrates the SLO-Scaler architecture. The system operates as a custom Kubernetes controller collecting metrics from Prometheus (CPU, memory, request rate) and distributed traces from Jaeger (per-service latency, call graph). These inputs feed three core modules: (1) a Bayesian LSTM predictor for workload, tail latency, and SLO violation probability; (2) a bottleneck localizer that walks the service dependency graph; and (3) an uncertainty-aware decision engine that issues scaling commands through the Kubernetes Scale API.

To keep the controller's coordination overhead bounded, SLO-Scaler adopts a communication-efficient distributed inference pattern from decentralized LLM serving over low-bandwidth nodes [36]. Each service-side monitor performs local aggregation over the lookback window and transmits compact predictive summaries (the posterior mean, variance, violation probability, and anomaly score) rather than forwarding every raw sample to the controller. The controller can therefore combine asynchronously arriving summaries while retaining the confidence information required by the scaling policy. This separation between local summarization and global SLO coordination reduces telemetry traffic and limits sensitivity to delayed updates without changing the bottleneck-localization or confidence-gating logic.

### B. Bayesian LSTM Prediction Module

Given per-service metrics $\mathbf{x}_t = \left[r_t, l_t^{p95}, c_t, m_t\right]$ (request rate, p95 latency, CPU, and memory utilization) over a lookback window of $W$ steps, we predict future values at horizon $H$ using a two-layer LSTM with Monte Carlo (MC) Dropout as a Bayesian approximation. At inference, $K$ stochastic forward passes yield predictions $\{\hat{\mathbf{x}}_{t+H}^{(k)}\}_{k=1}^{K}$, from which we compute the predictive mean and variance:

$$\hat{\mu}_{t+H} = \frac{1}{K}\sum_{k=1}^{K} \hat{\mathbf{x}}_{t+H}^{(k)} \quad (1)$$

$$\hat{\sigma}_{t+H}^{2} = \frac{1}{K}\sum_{k=1}^{K} \left(\hat{\mathbf{x}}_{t+H}^{(k)} - \hat{\mu}_{t+H}\right)^{2} \quad (2)$$

The SLO violation probability for service $s$ is the fraction of MC samples whose predicted p95 latency exceeds the SLO target $\tau_s$:

$$P_{\text{viol}}^{(s)} = \frac{1}{K}\sum_{k=1}^{K} \mathbf{1}\left[\hat{l}_{t+H}^{p95,(k)} > \tau_s\right] \quad (3)$$

where $\mathbf{1}[\cdot]$ is the indicator function. The confidence interval for $P^{(s)}_{\text{viol}}$ is derived from a Beta posterior with $\alpha = n_{\text{viol}} + 1$ and $\beta = K - n_{\text{viol}} + 1$.

### C. Bottleneck Localization via Dependency Graph

The service dependency graph $G = (V, E)$ is constructed from Jaeger traces, where vertices represent microservices and directed edges indicate call relationships. We compute an anomaly score per service:

$$a_s = \frac{l_s^{p95} - \bar{l}_s}{\max\left(\sigma_{l_s}, \epsilon_l\right)} \tag{4}$$

where $\bar{l}_s$ and $\sigma_{l_s}$ are the rolling mean and standard deviation of $s$'s p95 latency, and $\epsilon_l = 1$ ms prevents divergence. A weighted random walk on $G$ starts from the entry service with transition probabilities:

$$P(u \to v) = \frac{\max\left(a_v, 0\right) + \epsilon}{\sum_{w \in \text{children}(u)} \left(\max\left(a_w, 0\right) + \epsilon\right)} \tag{5}$$

where $\epsilon = 0.01$ ensures smoothing when all scores are non-positive. After $N$ random walks, the most-visited service is identified as the bottleneck. This approach extends the TopoRank method by incorporating predicted anomaly scores from our Bayesian LSTM for proactive identification.

### D. Uncertainty-Aware Scaling Decision Engine

Let $[P_{\text{lo}}, P_{\text{hi}}]$ denote the 90% credible interval of $P^{(s)}_{\text{viol}}$. The engine uses a four-case policy:

$$\text{action}(s) = \begin{cases} \text{scale-up} & \text{if } P_{\text{lo}} > \theta_{\text{hi}},\ s \in \mathcal{B} \\ \text{cautious-up} & \text{if } P_{\text{lo}} \le \theta_{\text{hi}},\ P_{\text{hi}} \ge \theta_{\text{lo}},\ s \in \mathcal{B} \\ \text{scale-down} & \text{if } P_{\text{hi}} < \theta_{\text{lo}} \text{ for } M \text{ consec.} \\ \text{hold} & \text{otherwise} \end{cases} \tag{6}$$

where $\theta_{\text{hi}} = 0.6$ and $\theta_{\text{lo}} = 0.2$ are thresholds, $\mathcal{B}$ is the set of bottleneck services, and $M = 3$ is the consecutive-window requirement for scale-down. Services not in $\mathcal{B}$ default to *hold*, avoiding unnecessary scaling. The *cautious-up* action adds a single replica, preventing aggressive scaling under high uncertainty. A cooldown of $T_{\text{cd}} = 60$s suppresses oscillation.

The scale-up target replica count incorporates uncertainty via the coefficient of variation:

$$R_s^* = \left\lceil R_s^{\text{cur}} \cdot \frac{\widehat{\mu}_{r,t+H}}{\max\left(\widehat{\mu}_{r,t}, \epsilon_r\right)} \cdot \left(1 + \gamma \cdot \frac{\widehat{\sigma}_{r,t+H}}{\max\left(\widehat{\mu}_{r,t+H}, \epsilon_r\right)}\right)\right\rceil \tag{7}$$

where $\gamma = 0.5$ is the safety margin and $\epsilon_r = 1$ RPS prevents division by zero. Scale-down removes one replica at a time: $R_s^* = \max\left(R_s^{\min}, R_s^{\text{cur}} - 1\right)$.

## IV. Experimental Setup

### A. Testbed and Workload

We deploy the DeathStarBench Social Network on a Kubernetes cluster with 5 worker nodes (8 vCPUs, 16 GB RAM each). The application comprises 12 microservices monitored by Prometheus (5 s interval) and Jaeger. The SLO target is $\tau = 200$ ms for end-to-end p95 latency. Locust generates HTTP traffic with profiles derived from the Alibaba 2022 trace [37] under three patterns: **Periodic** (sinusoidal, 120–280 RPS), **Bursty** (periodic baseline with a sustained 3× burst from minute 30 to 44), and **Long-tail** (periodic with sporadic 5× micro-bursts). Each experiment runs for 60 minutes with a 10-minute warm-up.

### B. Baselines and Metrics

We compare against: **K8s HPA** (CPU target 60%), **KEDA** (Prometheus HTTP rate, 100 RPS/pod), and **LSTM-Pred** (point-forecast LSTM scaling 5 minutes ahead). Metrics include SLO violation rate, p95/p99 latency, average replicas, scaling events, CPU-seconds, and cold-start recovery time. The Bayesian LSTM uses $W = 20$ steps, $H = 5$ steps, $K = 30$ MC samples, and a dropout rate of $p = 0.1$, pre-trained on 4 hours of baseline data.

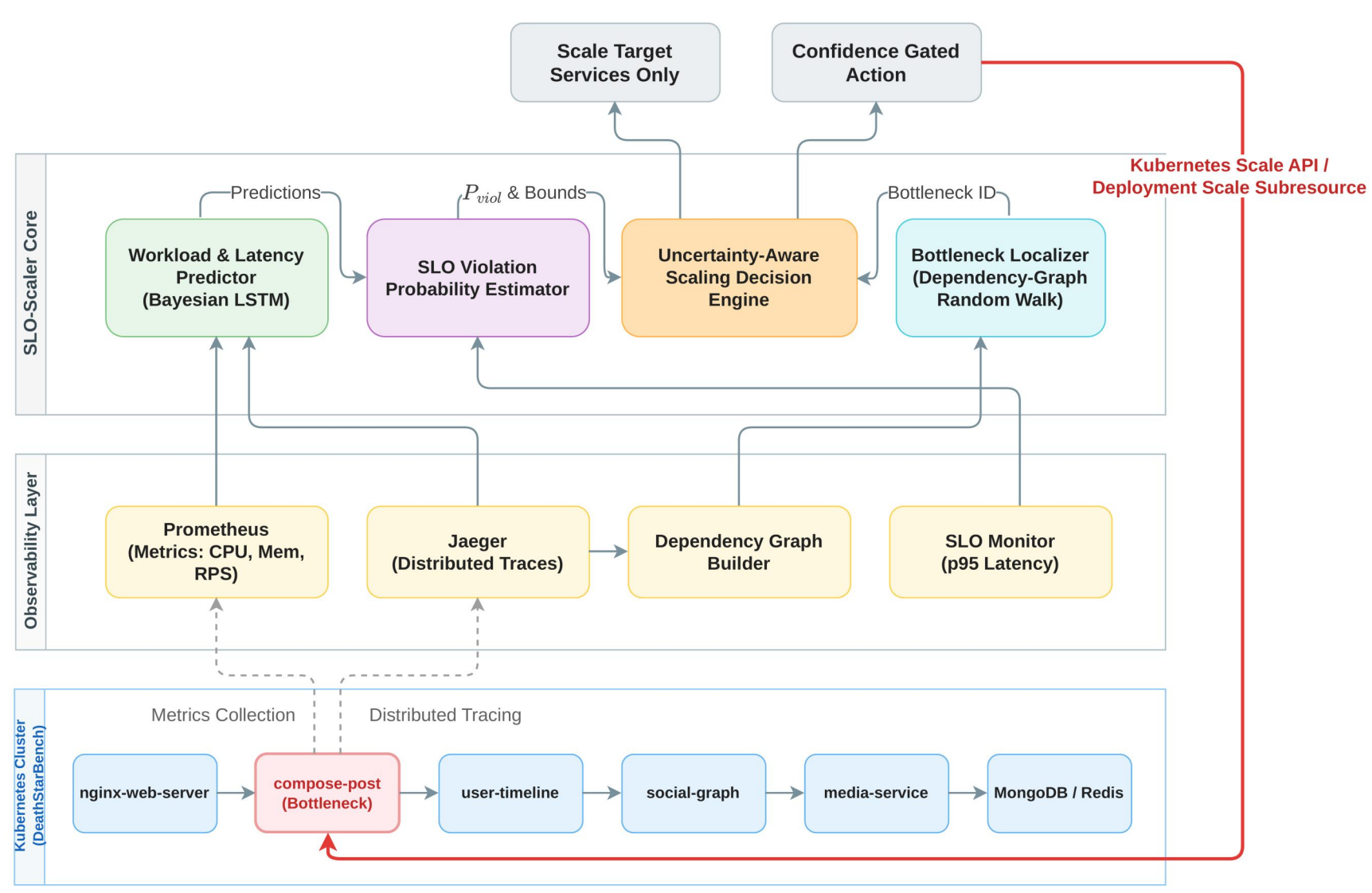

Fig. 1. Architecture of the SLO-Scaler framework deployed on DeathStarBench Social Network. The observability layer collects metrics from Prometheus and traces from Jaeger. The core predicts violation probability $P_{\text{viol}}$ via the Bayesian LSTM, localizes bottleneck services through dependency-graph random walks, and issues confidence-gated scaling commands via the Kubernetes Scale API.

TABLE I.
PERFORMANCE COMPARISON UNDER BURSTY TRAFFIC. BEST IN **BOLD**.

| Metric | HPA | KEDA | LSTM-P | Ours |
|---|---|---|---|---|
| SLO Viol. Rate (%) | 15.3 | 11.8 | 9.4 | **6.7** |
| p95 Latency (ms) | 218 | 196 | 183 | **162** |
| p99 Latency (ms) | 347 | 298 | 271 | **238** |
| Avg. Replicas | 4.8 | 4.2 | 3.9 | **3.2** |
| Scaling Events | 87 | 72 | 58 | **36** |
| CPU-sec ($\times 10^3$) | 42.1 | 37.6 | 34.8 | **28.3** |
| Recovery Time (s) | 48 | 35 | 22 | **16** |

TABLE II.
SLO VIOLATION RATE (%) AND AVERAGE REPLICAS ACROSS WORKLOAD PATTERNS.

| Pattern | HPA | KEDA | LSTM-P | Ours |
|---|---|---|---|---|
| *SLO Violation Rate (%)* | | | | |
| Periodic | 5.2 | 4.1 | 3.5 | **2.1** |
| Bursty | 15.3 | 11.8 | 9.4 | **6.7** |
| Long-tail | 11.7 | 8.9 | 7.6 | **5.3** |
| *Avg. Replicas* | | | | |
| Periodic | 3.6 | 3.2 | 3.0 | **2.5** |
| Bursty | 4.8 | 4.2 | 3.9 | **3.2** |
| Long-tail | 4.3 | 3.8 | 3.5 | **2.9** |

## V. RESULTS AND ANALYSIS

### A. *Overall Performance*

Table I summarizes the results under bursty traffic. SLO-Scaler achieves the lowest SLO violation rate of 6.7%, representing reductions of 56%, 43%, and 29% relative to HPA, KEDA, and LSTM-Pred, respectively. Average replicas decrease by 18–33% and scaling events by 38–59%, demonstrating that targeted bottleneck-aware scaling is more resource-efficient than uniform scaling. Consistent improvements are observed under periodic and long-tail patterns (Table II).

### B. *Temporal Behavior Under Bursty Traffic*

Fig. 2 shows the p95 latency and cumulative SLO violation rate from a representative 60-minute experiment. K8s HPA exhibits the most pronounced latency spikes, with peak p95 above 300 ms and approximately 48 s recovery time. KEDA responds faster via HTTP-rate metrics but still exceeds the SLO threshold during burst onset. LSTM-Pred begins pre-scaling about 60 s before the burst, reducing peak latency, but occasionally over-scales during recovery due to the absence of uncertainty quantification.

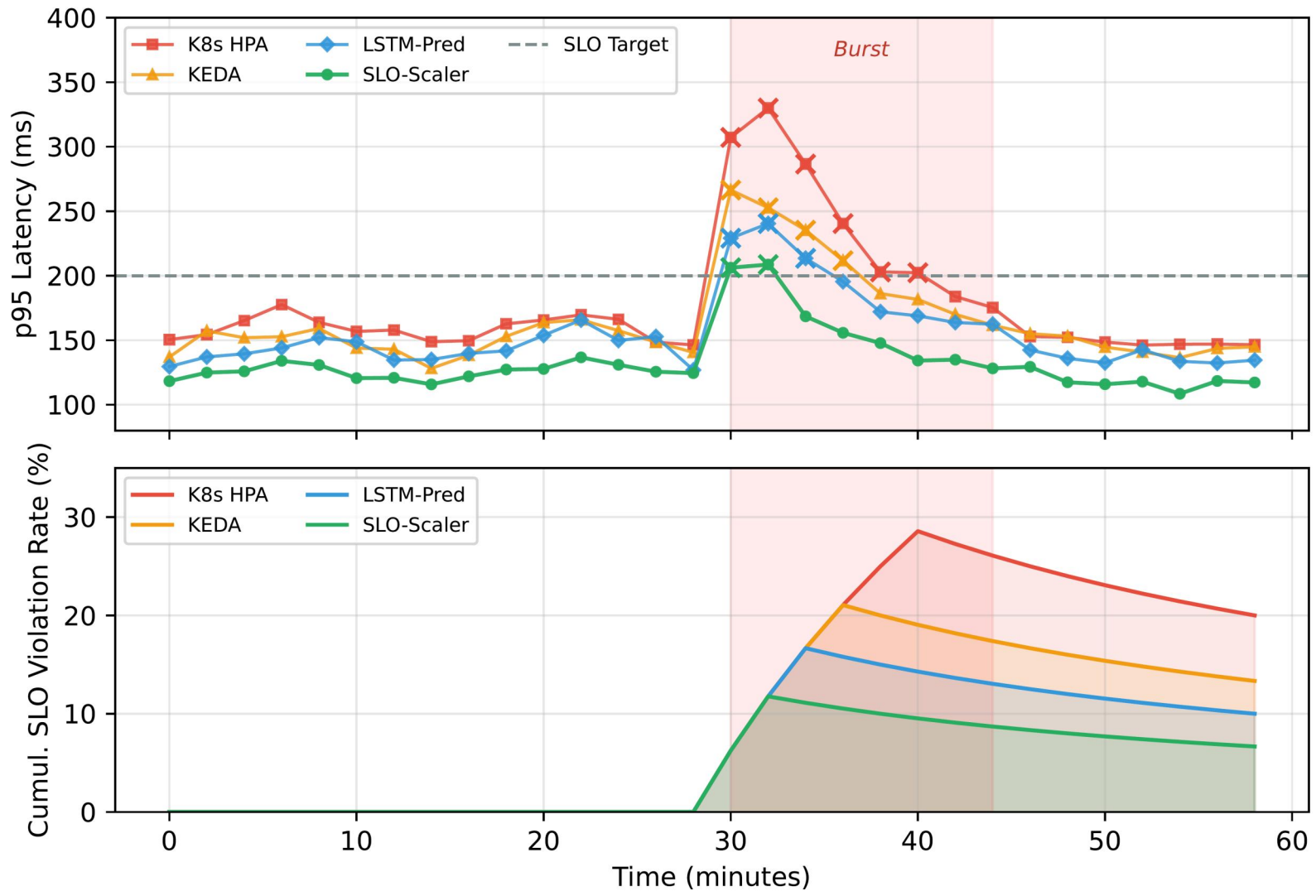

Fig. 2. P95 latency (top) and cumulative SLO violation rate (bottom) under bursty traffic. Shaded region: burst period (minutes 30–44). Table 1 reports averages over repeated runs.

SLO-Scaler initiates cautious pre-scaling when the violation probability begins rising and scales aggressively only when the lower confidence bound exceeds $\theta_{\text{hi}}$. This staged response keeps p95 latency below the SLO target for most of the burst, with exceedances limited to the initial transient.

### C. *Bottleneck-Targeted Scaling*

Fig. 3 presents per-service scaling commands across methods. Each system-level scaling event in Table I may issue commands to multiple services; therefore, per-service command counts are slightly higher than event counts. K8s HPA and KEDA distribute commands nearly uniformly across all services, whereas SLO-Scaler concentrates 22 of its 39 commands on the identified bottleneck compose-post, with only 3–4 commands for non-bottleneck services. This targeted approach avoids unnecessary replica provisioning for services not on the critical path.

### D. *Ablation Study*

Table III evaluates two ablated variants under bursty traffic. Removing the bottleneck localizer increases average replicas by 28% and scaling events by 44%, as the system scales all services indiscriminately. Removing uncertainty quantification increases the violation rate to 8.5% and scaling events to 61, because the scaler reacts to every point prediction exceeding the threshold, triggering oscillations. Both components provide complementary benefits.

TABLE III.
ABLATION STUDY RESULTS UNDER BURSTY TRAFFIC.

| Variant | Viol. (%) | Avg. Rep. | Events |
|---|---|---|---|
| Full SLO-Scaler | **6.7** | **3.2** | **36** |
| w/o Bottleneck | 7.9 | 4.1 | 52 |
| w/o Uncertainty | 8.5 | 3.5 | 61 |

## VI. Conclusion

We presented SLO-Scaler, an uncertainty-aware autoscaling framework for microservices. By integrating Bayesian LSTM-based SLO violation prediction with uncertainty-gated scaling and bottleneck-aware dependency graph analysis, SLO-Scaler reduces SLO violations by 29–56%, scaling events by 38–59%, and average replicas by 18–33% compared to three baselines under bursty traffic, with consistent improvements under periodic and long-tail patterns. Several limitations merit discussion: the Bayesian LSTM requires approximately 4 hours of warm-up data; the MC Dropout approximation primarily captures epistemic uncertainty and may underestimate total uncertainty in non-stationary environments; and the bottleneck localizer assumes a relatively stable dependency graph. Future work will explore ensemble methods for improved uncertainty calibration, extend the framework to vertical scaling, and evaluate on larger production traces [38].

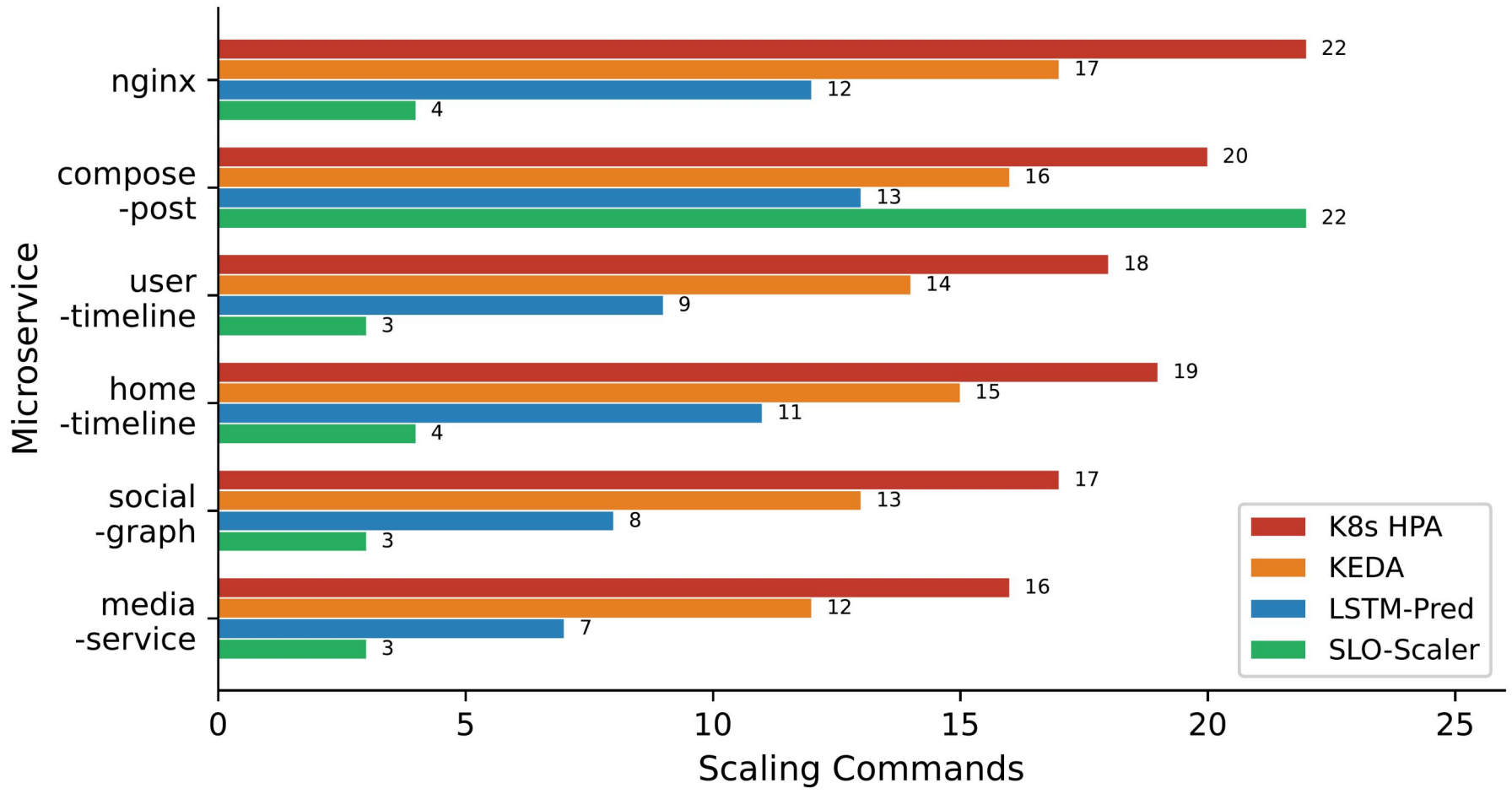

Fig. 3. Per-service scaling commands. SLO-Scaler concentrates commands on the bottleneck (compose-post), while baselines distribute them uniformly.

## References

[1] Y. Gan *et al.*, "An open-source benchmark suite for microservices and their hardware-software implications for cloud and edge systems," in *Proc. Int. Conf. on architectural support for programming languages and operating systems (ASPLOS)*, ACM, 2019, pp. 3–18.

[2] S. Xie, J. Wang, B. Li, Z. Zhang, D. Li, and P. C. K. Hung, "PBScaler: A bottleneck-aware autoscaling framework for microservice-based applications," *IEEE Transactions on Services Computing*, vol. 17, no. 2, pp. 604–616, 2024.

[3] S. Luo *et al.*, "An in-depth study of microservice call graph and runtime performance," *IEEE Transactions on Parallel and Distributed Systems*, vol. 33, no. 12, pp. 3901–3914, 2022.

[4] H. Qian, Q. Wen, L. Sun, J. Gu, Q. Niu, and Z. Tang, "RobustScaler: QoS-aware autoscaling for complex workloads," in *Proc. 38th IEEE int. Conf. on data engineering (ICDE)*, IEEE, 2022, pp. 2762–2775.

[5] B. Choi, J. Park, C. Lee, and D. Han, "pHPA: A proactive autoscaling framework for microservice chain," in *Proc. 5th Asia-Pacific workshop on networking (APNet)*, ACM, 2021, pp. 65–71.

[6] H. X. Nguyen, S. Zhu, and M. Liu, "Graph-PHPA: Graph-based proactive horizontal pod autoscaling for microservices using LSTM-GNN," in *Proc. 11th IEEE int. Conf. on cloud networking (CloudNet)*, IEEE, 2022, pp. 237–241.

[7] Y. Zhang, W. Hua, Z. Zhou, G. E. Suh, and C. Delimitrou, "Sinan: ML-based and QoS-aware resource management for cloud microservices," in *Proc. Int. Conf. on architectural support for programming languages and operating systems (ASPLOS)*, ACM, 2021, pp. 167–181.

[8] G. Yu, P. Chen, and Z. Zheng, "MicroScaler: Automatic scaling for microservices with an online learning approach," in *Proc. IEEE int. Conf. on web services (ICWS)*, IEEE, 2019, pp. 68–75.

[9] A. Abdel Khaleq and I. Ra, "Intelligent microservices autoscaling module using reinforcement learning," *Cluster Computing*, vol. 26, no. 5, pp. 2789–2800, 2023.

[10] K. Hu, L. Wen, M. Xu, and K. Ye, "MSARS: A meta-learning and reinforcement learning framework for SLO resource allocation and adaptive scaling for microservices," in *Proc. IEEE int. Symp. on parallel and distributed processing with applications (ISPA)*, IEEE, 2024, pp. 590–599.

[11] C. Meng, J. Tong, M. Pan, and Y. Yu, "HRA: An intelligent holistic resource autoscaling framework for multi-service applications," in *Proc. IEEE int. Conf. on web services (ICWS)*, IEEE, 2022, pp. 129–139.

[12] Z. Wang *et al.*, "DeepScaling: Autoscaling microservices with stable CPU utilization for large scale production cloud systems," *IEEE/ACM Transactions on Networking*, vol. 32, no. 5, pp. 3961–3976, 2024.

[13] A. Rossi, A. Visentin, D. Carraro, S. D. Prestwich, and K. N. Brown, "Forecasting workload in cloud computing: Towards uncertainty-aware predictions and transfer learning," *Cluster Computing*, vol. 28, no. 4, 2025.

[14] K. Yang, F. Wang, and R. Jiang, "SLO-Aware Graph Forecasting for Intelligent Autoscaling in API and Microservice Backends," 2026, DOI: 10.13140/RG.2.2.28071.92322.

[15] J. Qiu, J. Sun, Y. Xue, Y. Wang, Z. Yang, and C. Zhang, "Attention-Enhanced Long Sequence Temporal Modeling Method for Backend CPU Load Prediction Tasks," 2026.

[16] S. Zhang, S. Shawulieti, and S. Pi, "Cold-Start Budgeting for Serverless ML Inference via Workload Forecasting and Model State Tiering," in *Proc. 3rd Int. Conf. on Image Processing and Artificial Intelligence (ICIPAI)*, IEEE, 2026, pp. 345-351.

[17] D. Huang, N. Zhao, W. Liu, and N. Sang, "Target-Aware Augmentation for Rare-Event Prediction under Tabular Covariate Shift," 2026, DOI: 10.13140/RG.2.2.32056.51202.

[18] Y. Nie, J. Wang, R. Yan, Y. Wang, Z. Ma, and Y. Wu, "Graph-based financial fraud detection with calibrated risk scoring and structural regularization," in *Proc. Int. Conf. on AI Decision-Making and Management*, 2026, pp. 169-176.

[19] Y. Ma et al., "Research on fine-tuning algorithms for Large Language Models integrating Uncertainty Modeling and External Memory Augmentation," *PLoS One*, vol. 21, no. 6, Art. no. e0351493, 2026.

[20] X. Huang, T. Xia, Y. Zhou, J. Liao, L. Ren, and F. Chang, "Risk-Calibrated Uncertainty-Aware Edge-Cloud Scheduling for Reliable Multi-Tenant ML Inference," in *Proc. 6th Int. Conf. on Electronics, Circuits and Information Engineering (ECIE)*, IEEE, 2026, pp. 573-578.

[21] X. Meng, L. Zheng, Y. Duan, S. Hu, and J. Chen, "Dynamic Enterprise Audit Risk Identification Through Graph-Temporal

Transformer Representation Learning," *ResearchGate Preprint*, 2026, DOI: 10.13140/RG.2.2.26915.21280.

[22] J. Zhang, B. Chen, Y. Tang, F. Chen, Y. Zhan, and Z. Hu, "Drift-Aware Straggler Mitigation for Heterogeneous Split Federated Learning via Proactive Cut-Layer and Asynchronous Server Updates," 2026, DOI: 10.13140/RG.2.2.26441.10080.

[23] J. Jiang, J. Hu, and Y. Lyu, "Cost-Aware Cross-Cloud Federated Learning with Learned Client Routing and Aggregation Selection," in *Proc. 8th Int. Conf. on Internet of Things, Automation and Artificial Intelligence (IoTAAI)*, IEEE, 2026, pp. 111-116.

[24] T. Xia, X. Huang, Y. Zhou, F. Chang, and L. Ren, "Coordinating Training and Inference in Heterogeneous Cloud GPU Clusters with Learned Interference Models," in *Proc. 8th Int. Conf. on Internet of Things, Automation and Artificial Intelligence (IoTAAI)*, IEEE, 2026, pp. 104-110.

[25] C. He, Y. Gong, Y. Ma, B. Zhang, and S. Wang, "Subspace-Deconvolution Parameterization for Efficient Large Language Model Adaptation," 2026, DOI: 10.13140/RG.2.2.34546.88008.

[26] Z. Xiao, Q. Guo, Y. Wang, B. Wang, and H. Lu, "AgentDropout: Dynamic Redundancy Elimination for Multi-Agent Collaboration Efficiency," 2026, DOI: 10.13140/RG.2.2.15489.01125.

[27] Z. Liu, K. Wu, S. Dang, and Y. Yang, "A Retrieval-Augmented Log-Metric-Trace Twin for Backend Failure Diagnosis," 2026, DOI: 10.13140/RG.2.2.34050.64969.

[28] W. Ma, Z. Liu, R. Jiang, and Y. Jiang, "TopoRCA-Lite: Dynamic Topology-Aware Root Cause Localization under Limited Observability," in *Proc. 6th Int. Conf. on Machine Learning and Intelligent Systems Engineering (MLISE)*, IEEE, 2026, pp. 676-681.

[29] R. Hu, Y. Zheng, R. Fang, and J. Zhou, "VeriTrail-RCA: Evidence-Verified Root Cause Localization for Microservice Backends," in *Proc. 6th Int. Conf. on Machine Learning and Intelligent Systems Engineering (MLISE)*, IEEE, 2026, pp. 670-675.

[30] Y. Zheng, J. Zhou, R. Hu, and R. Fang, "Evidence-Verified LLM Agents for Safe Backend Incident Remediation," in *Proc. 3rd Int. Conf. on Image Processing and Artificial Intelligence (ICIPAI)*, IEEE, 2026, pp. 321-326.

[31] C. Zhang, J. Sun, Z. Hu, J. Zhang, Y. Wang, and Z. Yang, "VeriAct-Agent: Evidence-Verified Tool Execution for Reliable LLM Agents," in *Proc. 6th Int. Conf. on Machine Learning and Intelligent Systems Engineering (MLISE)*, IEEE, 2026, pp. 240-245.

[32] L. Tang, Q. Cheng, Q. Guo, and Z. Ke, "Lineage-Aware Semantic Validation and Verified Self-Repair for Reliable Data Pipelines," 2026, DOI: 10.13140/RG.2.2.18743.89763.

[33] L. Tang, Y. Mei, X. Ren, Y. Ke, Z. Ke, and Z. Bian, "DriftMender: LLM-Guided Semantic Repair for Schema Drift in Real-Time ETL Pipelines," 2026, DOI: 10.13140/RG.2.2.19248.52483.

[34] Z. Zhu, R. Fang, and R. Hu, "Experience Retrieval Compression for Continual Adaptation of LLM Agents," 2026, DOI: 10.13140/RG.2.2.32706.34245.

[35] J. Y. Su and X. Yan, "Process-Aware Multimodal Drawing Analytics for AI-Assisted Expressive Arts Support in Educational Settings," 2026, DOI: 10.13140/RG.2.2.23865.53608.

[36] Z. Wang, S. Shu, Y. Wang, I. H. Lai, and C. C. Peng, "Communication-Efficient Decentralized LLM Inference over Low-Bandwidth Distributed Nodes," 2026, DOI: 10.13140/RG.2.2.34464.96000.

[37] Alibaba Group, "Cluster trace: Microservices v2022." https://github.com/alibaba/clusterdata/tree/master/cluster-trace-microservices-v2022, 2022.

[38] A. Detti, L. Funari, and L. Petrucci, "Bench: An open-source factory of benchmark microservice applications," *IEEE Transactions on Parallel and Distributed Systems*, vol. 34, no. 3, pp. 968–980, 2023.